\documentclass[sigconf]{acmart}
\AtBeginDocument{%
  }

\setcopyright{acmlicensed}
\copyrightyear{2027}
\acmYear{2027}
\acmConference[WebSci '27]{ACM Web Science Conference}{2027}{TBD}
\acmDOI{10.1145/XXXXXXX.XXXXXXX}
\acmISBN{978-1-4503-XXXX-X/27/XX}

\usepackage{multirow}
\begin{document}

\title{
Information Operations Exploit APIs to Manipulate Social Media 
}

\author{Manita Pote}
\correspondingauthor
\affiliation{%
  \institution{Indiana University}
  \city{Bloomington}
  \state{Indiana}
  \country{USA}
}

\author{Alessandro Flammini}
\affiliation{%
  \institution{Indiana University}
  \city{Bloomington}
  \state{Indiana}
  \country{USA}
}

\author{Filippo Menczer}
\affiliation{%
  \institution{Indiana University}
  \city{Bloomington}
  \state{Indiana}
  \country{USA}
}

\begin{abstract}
Research on information operations has focused primarily on the coordinated accounts involved in such campaigns and in the content they promote.  
Bad actors manage the activity of inauthentic accounts through programmatic interfaces (APIs), but the role of this underlying infrastructure in enabling and uncovering such coordinated behaviors remains largely unexplored. 
This study investigates how the Twitter (now X) API was used to coordinate influence operations on the platform. 
Since API access is managed through developer apps, we extract app metadata from 43 operations taken down between 2018 and 2021 and apply three complementary computational approaches to study diverse forms of manipulation.
Our analysis reveals several recurring patterns: API-as-a-service infrastructure reused across campaigns, gaming apps performing unauthorized engagement without user consent, spoofed app names to hide activity origins, and synchronized app networks indicating centralized control. 
Adding app usage as a coordination indicator provides a modest but consistent improvement in detection. 
These findings suggest that social media APIs play an important role in enabling online manipulation, and that the removal of app-level metadata by Twitter/X has eliminated an important tool for platform transparency and moderation.
\end{abstract}




\maketitle

\section{Introduction}

Influence campaigns, also known as information operations (IOs)~\cite{James2022}, are efforts to manipulate social media platforms with the aim of influencing public opinion. 
Researchers have extensively studied how these campaigns unfold on social media, such as characterizing the actors involved~\cite{senateIntelligenceCommittee2019}, identifying common tactics~\cite{stewart2018examining, zannettou2020characterizing}, developing detection methods~\cite{pacheco2021uncovering, luceri2024}, and assessing their impacts~\cite{eady2023exposure}. 
Despite this growing body of work, one underexplored aspect of information operations is the role of application programming interfaces (APIs), which allow programmatic access to social media platforms. 

Developer apps are mechanisms provided by platforms to manage public API access. 
Users create apps to obtain authorization keys for a platform. 
These keys are in turn used by code that interacts with the platform to post content, gather data, or provide third-party services.  
These services can extend the core functionality of a social media platform and act as intermediaries between users and the platform itself. 
When granted authorization, apps can take actions on behalf of users and access sensitive data through mechanisms such as delegation or access tokens. 

While most apps are benign, privileged access to platform data via APIs introduces substantial potential for misuse. 
A well-known example is the Cambridge Analytica incident, where data harvested via a third-party Facebook app was exploited for targeted political advertising~\cite{cnbc2018cambridge}.
While this case highlights how data access can be abused, the use of APIs for manipulation, for example as part of information operations, remains an under-examined threat.  

Prior research has explored the use of automation apps and spoofed versions of popular applications in information operations~\cite{saeed2024unravel} as well as the use of APIs to manage networks of social bots and other inauthentic accounts~\cite{bradshaw2017troops}. 
We extend this research by shifting the unit of analysis from accounts to apps, investigating how third-party applications facilitate coordinated inauthentic behavior. 
Specifically, we ask: 

\begin{itemize}
    \item \textbf{RQ1:} What forms of manipulation are observed through third-party app usage patterns?
    \item \textbf{RQ2:} Does incorporating app usage as a coordination indicator improve detection performance?
\end{itemize}

To answer RQ1, we present three case studies that analyze the same app usage data using complementary computational methods, each identifying groups of potentially coordinated apps or accounts. 
We then apply qualitative analysis to understand the forms of manipulation enabled by app usage. 
As analyzing all groups individually is impractical, we restrict our analysis to the ten largest groups in each case and report the common behavioral patterns observed across them. 
Qualitative analysis is essential here because close examination of what those accounts are actually doing reveals whether coordination constitutes manipulation, what form it takes, and which behavioral signals detection methods can leverage to model it.

For RQ2, we add app usage as a coordination indicator into an unsupervised detection method to evaluate whether it improves detection performance. 

This study makes the following contributions:
\begin{itemize}
    \item We provide evidence of shared API-as-a-service infrastructures, gaming apps that perform unauthorized engagement actions without user knowledge, and spoofed app names designed to obscure the origin of activity.
    \item We find that apps sharing overlapping user accounts expose centrally controlled amplification networks, where the same pool of accounts posts through multiple apps to boost content at scale. 
    \item We show that accounts sharing similar app usage patterns coordinate through distinct tactics, including synchronized social media management tools, custom fake news infrastructure, and RSS-based propaganda automation.
    \item We demonstrate that adding app usage as a coordination indicator provides a modest but consistent improvement in detection.
\end{itemize}

These findings demonstrate that identifying manipulation requires examining not only suspicious accounts and content, but also the third-party applications through which activity originates. 
This emphasizes the importance of incorporating app-level signals into comprehensive strategies for curbing manipulation. 
Twitter/X has not only made it extremely expensive for researchers to access its API, but also removed access to critical app metadata, making it impossible to conduct these kinds of analysis and therefore more difficult to detect coordinated manipulation campaigns.

\section{Data}

We use two datasets in this study: a broad collection of Twitter information operation campaigns to characterize third-party app usage patterns (RQ1), and a labeled subset of these campaigns with matched organic control accounts to evaluate coordination detection performance (RQ2).

\subsection{App usage dataset (RQ1)}

We analyze app meta-data from 43 IOs taken down between 2018 and 2021 by the Twitter Moderation Research Consortium. 
These datasets consist of archives of accounts suspended by Twitter for suspected involvement in information operations across various countries. 
Each dataset includes each profile's complete tweet history since the time it was created, along with the associated profile metadata. 
This data is no longer available from Twitter.\footnote{\url{web.archive.org/web/20240913160929/https://transparency.x.com/en/reports/moderation-research}}

Our analysis focuses on the \texttt{source} field within each tweet object, which indicates the client application used to post the tweet. 
For clarity, we refer to the source as `app' throughout this paper.
Across all datasets, we identify 3,877 distinct apps used by IO accounts. 
We apply a set of filters to focus on campaign-specific apps rather than general-purpose tools. 
First, we remove accounts with fewer than 10 total tweets to retain only active users. 
Second, we require each app to have at least 100 total tweets and at least 5 distinct users to filter out hobby or personal apps. 
Third, we cap the maximum number of users per app at 500. This excludes broadly used apps, such as official platform clients, that are unlikely to reflect campaign-specific coordination tool. 
We compile all IO activity generated by 496 apps that pass the above filtering process. 
A summary of the dataset is presented in Table~\ref{tab:rq1_dataset_summary}.

\begin{table}
    \centering
    \caption{Summary of the dataset showing total IO activity generated by third-party applications. Original tweets include posts and replies.}
    \begin{tabular}{lr}
    \hline
    \bf Data & \bf Total \\ 
    \hline
        \textbf{Apps}            & 496 \\
        \textbf{Accounts}        & 8,597 \\
        \textbf{Original Tweets} & 16,442,468\\
        \textbf{Retweets}        & 4,631,582 \\
        \textbf{Total Tweets}    & 21,074,050 \\
        \hline
    \end{tabular}
    \label{tab:rq1_dataset_summary}
\end{table}

We acknowledge that the filter thresholds above are chosen heuristically. 
To assess sensitivity to these choices, we vary one filter parameter at a time while holding the others fixed and measure the overlap in detected apps or accounts against the reference configuration.
These results are reported within the respective case study sections.

\subsection{Labeled dataset (RQ2)}
\label{sec:labeled-datasets}

\begin{table}
    \centering
    \small
    \setlength{\tabcolsep}{3pt}
    \renewcommand{\arraystretch}{1.1}
    \caption{Dataset summary for RQ2. Ratio = IO / Ctrl.}
    \label{tab:rq2_dataset}
    \begin{tabular}{lrrrrrr}
    \hline
    \textbf{Camp.} & \textbf{Tweets} & \textbf{RTs} & \textbf{Apps} & \textbf{IO} & \textbf{Ctrl} & \textbf{Ratio (\%)} \\
    \hline
    Russia\_1                  & 10.51M & 3.33M & 2,260 & 3,184 & 21,010 & 15.15 \\
    Russia\_2    & 1.01M & 712.7k & 262  & 275  & 1,427  & 19.27 \\
    Russia\_3          & 32.1k & 0.5k  & 30    & 4    & 555    & 0.72  \\
    Russia\_4          & 1.80M & 21.8k & 1,794 & 20   & 25,550 & 0.08  \\
    Russia\_5          & 0.69M & 3.7k  & 552   & 40   & 9,284  & 0.43  \\
    
    Armenia      & 139.5k & 0.5k & 172   & 26   & 1,104  & 2.36  \\
    Bangladesh & 60.7k & 1.0k & 141   & 9    & 543    & 1.66  \\
    Catalonia & 121.6k & 6.0k & 99    & 35   & 1,856  & 1.89  \\
    China\_1             & 3.62M & 356.0k & 1,497 & 587  & 28,448 & 2.06  \\
    China\_2             & 3.37M & 668.1k & 2,089 & 181  & 25,889 & 0.70  \\
    Cuba                 & 6.11M & 3.34M & 1,224 & 461  & 20,109 & 2.29  \\
    Ecuador   & 1.44M & 577.6k & 709   & 620  & 12,858 & 4.82  \\
    Egypt\_UAE & 229.0k & 71.9k & 152  & 224  & 242    & 92.56 \\
    Ghana\_Nigeria & 81.9k & 16.3k & 103  & 58   & 777    & 7.46  \\
    
    Iran\_1              & 1.29M & 231.9k & 510  & 595  & 3,292  & 18.07 \\
    Iran\_2              & 1.62M & 502.3k & 592   & 322  & 4,780  & 6.74  \\
    Iran\_3             & 2.05M & 1.61M & 220   & 198  & 1,534  & 12.91 \\
    Iran\_4              & 548.3k & 90.0k & 439  & 1,860 & 5,054  & 36.80 \\
    Iran\_5         & 43.4k & 0.6k  & 51    & 23   & 773    & 2.98  \\
    Iran\_6         & 1.27M & 100.4k & 593  & 188  & 11,535 & 1.63  \\
    
    Qatar        & 999.2k & 163.5k & 512 & 24   & 12,960 & 0.19  \\
    
    Spain     & 116.9k & 27.0k & 52   & 179  & 1,056  & 16.95 \\
    Thailand     & 140.1k & 10.8k & 43   & 208  & 1,926  & 10.80 \\
    UAE       & 1.69M & 644.0k & 559  & 3,349 & 6,638  & 50.45 \\
    Venezuela\_2         & 1.24M & 1.4k  & 241   & 501  & 4,316  & 11.61 \\
    Venezuela\_1         & 721.6k & 29.0k & 184  & 32   & 2,680  & 1.19  \\
    \hline
    \end{tabular}
\end{table}

To answer RQ2, we use 26 labeled datasets consisting of IO and control accounts, that were curated by \citet{seckin2025labeled} to benchmark detection performance. 
As for the app usage dataset, we applied some filtering to clean up the data. 
First, we discarded accounts with fewer than 10 total tweets. 
Second, leveraging prior work~\cite{pacheco2021uncovering}, we considered various indicators of coordinated behavior based on shared hashtags, shared URLs, retweeted users, and co-retweets. 
We retained only tweets that include indicators used by at two users at least 10 times in total. 
Table~\ref{tab:rq2_dataset} summarizes the resulting datasets. 

\section{Related work}

In this section, we review relevant literature on information operations as well as on the privacy and security risks associated with third-party applications.

\subsection{Information operations}

A crucial first step in curbing social media information operations is to understand how these campaigns operate. 
Prior research has extensively characterized the strategies and behaviors associated with such operations, especially ones originating from Russia. 
For instance, \citet{howard2018ira, zannettou2019disinformation, kristina2022spread} found that information campaigns are frequently used to spread disinformation. \citet{stewart2018examining} analyzed the behavior of Russian trolls around the \#BlackLivesMatter movement and found that these actors infiltrated both right- and left-leaning political communities. 
Similarly, \citet{dutt2018senator} studied Facebook ads from the 2016 Russian campaign and observed that the most effective ads were highly personalized and often carried a negative tone.
\citet{zannettou2020characterizing} highlighted the widespread dissemination of political imagery in Russian influence campaigns. 

\citet{astrofurfSchoch, keller2020political} demonstrated that coordinated activity is often used to simulate grassroots support, a tactic known as astroturfing. 
Other studies have identified the use of bots to amplify content and drive engagement~\cite{shao2018bot, geissler2023russianpropagandasocialmedia}.
\citet{Torres2022deletions} observed coordinated patterns of repeated likes and unlikes to manipulate platform ranking algorithms.
\citet{pote2025coordinated} reported on coordinated reply attacks targeting influential figures with the goal of either supporting or harassing them. 
Researchers have also documented evolving techniques in newer media ecosystems. 
\citet{luceri2025tiktok} found that AI-generated voice-overs and split-screen videos are used to amplify political narratives on TikTok. 
\citet{Yang2023Anatomy-AI-botnet} uncovered a large network of coordinated Twitter bots managed through ChatGPT, involved in cryptocurrency scams.  
Likewise, \citet{federico2025cross} found cross-platform coordination efforts involving the dissemination of links to suspicious websites and the promotion of specific accounts on X and YouTube.
Finally, and most related to this paper, \citet{saeed2024unravel} reported on the use of fake or spoofed versions of legitimate third-party applications to create a false sense of legitimacy.

\subsection{Privacy and security risks}

One of the earliest major incidents involving privacy breaches by third-party applications on Facebook was reported in 2010~\cite{Beaumont2010Facebook}. 
Apps such as Farmville and Mafia Wars were found to be illegally transmitting Facebook user IDs, names, and friend lists to advertising and tracking agencies. 
In 2013, Cambridge academic Aleksandr Kogan launched an app called \textit{ThisIsYourDigitalLife}, which collected data not only from its users but also from their friends, ultimately affecting millions without consent. 
This data was later exploited by Cambridge Analytica for targeted political advertising, raising global concerns about data misuse and election interference~\cite{cnbc2018cambridge}. 
Since then, a series of similar data misuse incidents have come to light, including the \textit{Rankwave} lawsuit filed by Facebook in 2019, which accused a South Korean analytics firm of unauthorized data collection through its apps~\cite{losingface, techcrunch2019rankwave}. 

These events have prompted numerous studies on privacy control and third-party app permissions in social media platforms. \citet{madejski2011failure} found that Facebook's access control settings were often misconfigured by users, resulting in unintended data exposure. 
\citet{Wang2011ThirdParty} analyzed 1,800 Facebook apps and found flaws in privacy consent dialogs, including poor layout, limited control, and unclear data access. 
They proposed redesigning the interface to improve user control and alert users to privacy violations.
\citet{Anthony2012Collaborative} proposed a Collaborative Privacy Management framework that acts as an intermediary between third-party apps and platforms.
It would let users approve specific permissions and share or adopt privacy settings to support informed, collective decision-making.

\section{Case studies}

While prior research has offered valuable insights into the risks of data misuse through third-party applications, their strategic use in information operations has not been fully examined yet. 
The present study focuses on the role of third-party apps as the infrastructure underlying IOs. 
We present three case studies examining IO manipulation tactics facilitated through third-party applications. 

\subsection{Case 1: App network based on name similarity}

In the first case study, we construct a network of apps based on the textual similarity between their names. 
The similarity between two app names, $n_i$ and $n_j$, is computed using the normalized longest prefix match:
\begin{equation}
    S(n_i, n_j) = \frac{L(n_i, n_j)}{\min(\text{length}(n_i), \text{length}(n_j))}
\end{equation}
where $L(n_i, n_j)$ denotes the length of the longest common prefix between $n_i$ and $n_j$, and $\text{length}(n_i)$ corresponds to the number of characters in a string $n_i$.
A higher $S(n_i, n_j)$ value indicates greater similarity between app names $n_i$ and $n_j$ ($0 \leq S(n_i, n_j) \leq 1$). 

We use this similarity measure to construct a network where nodes represent apps and the edge between two nodes is weighted by their app name similarity. 
The network is very dense and therefore we need to filter edges in order to identify meaningful clusters. 
To find a meaningful cutoff, we vary the edge similarity threshold and track two quantities: the number of connected components and the size of the largest component (Appendix~\ref{app:case1}, Figure~\ref{fig:percolation_analysis_case1}).
For low values of the threshold, all apps collapse into one large connected component.  
Past the 95th percentile, the network breaks apart into small tight groups, marking the percolation threshold where meaningful cluster structure emerges. 
We select the 99th percentile because it lies in the stable region after the transition, where the number of components and the size of the largest component have both stabilized. This corresponds to a similarity threshold of 0.48 and a network with 203 suspicious apps across 41 groups (Table~\ref{tab:network_stats} in Appendix), visualized in Figure~\ref{fig:networks}A. 

\begin{figure*}
    \centering
    \includegraphics[width=0.33\linewidth]{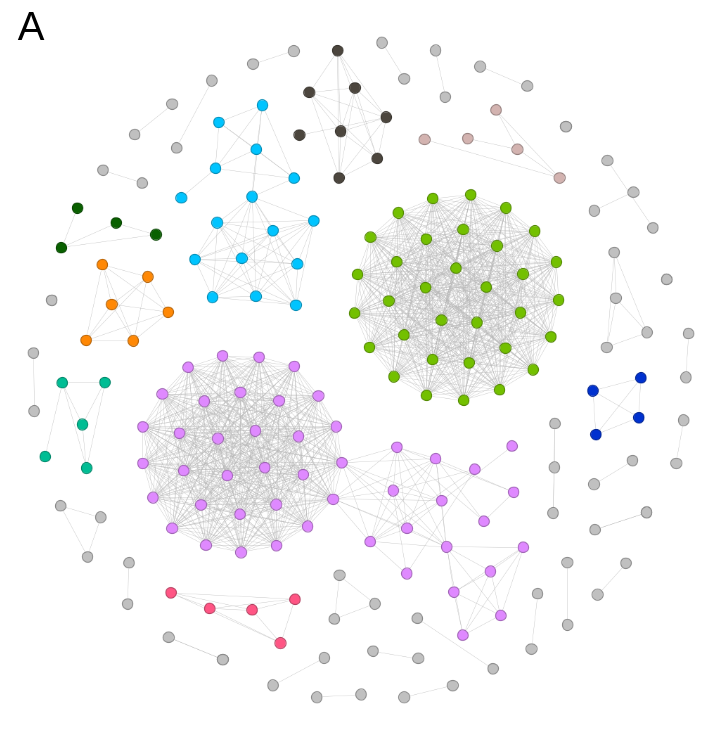}
    \includegraphics[width=0.33\linewidth]{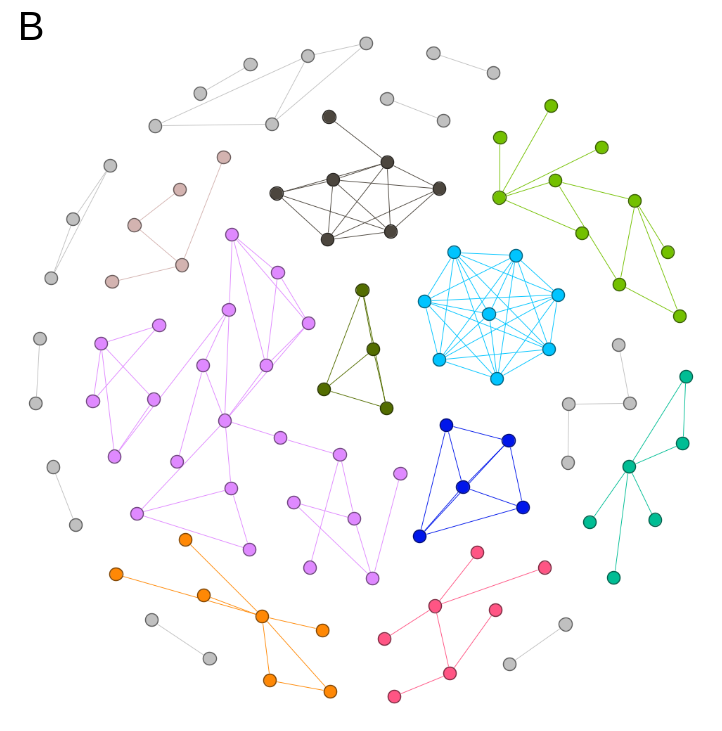}
    \includegraphics[width=0.33\linewidth]{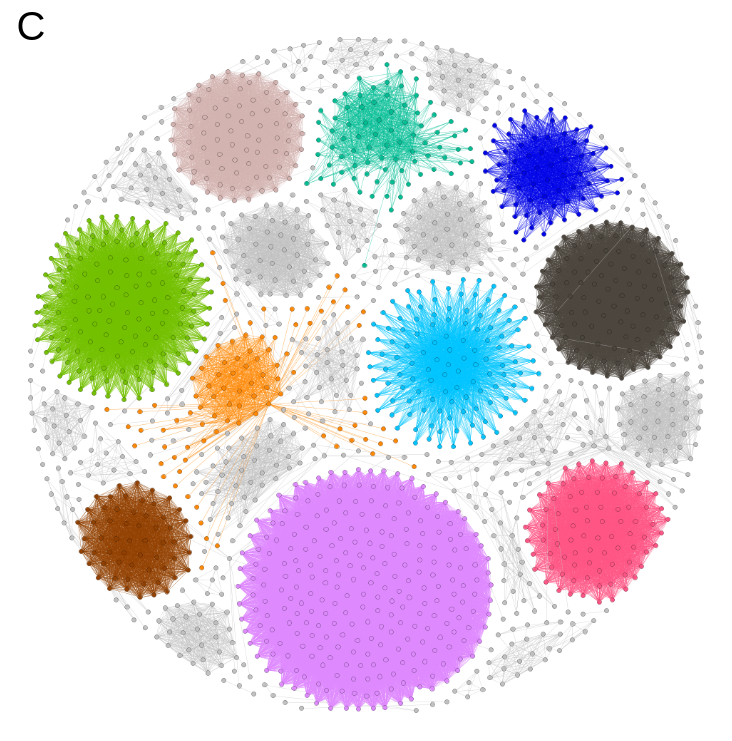}
    \caption{
    App and account coordination networks for the three case studies. (A)~App name similarity network (Case 1, 99th percentile threshold), (B)~account overlap network (Case 2, 99th percentile threshold), and (C)~account similarity network based on app usage patterns (Case 3, 99.9th percentile threshold). In each network, nodes represent apps (A, B) or accounts (C), and edges connect pairs above the percentile threshold. The ten largest components are highlighted in distinct colors.
    }
    \Description{App and account coordination networks for the three case studies. (A) App name similarity network (Case 1, 99th percentile threshold), (B) account overlap network (Case 2, 99th percentile threshold), and (C) account similarity network based on app usage patterns (Case 3, 99.9th percentile threshold). In each network, nodes represent apps (A, B) or accounts (C), and edges connect pairs above the percentile threshold. The ten largest components are highlighted in distinct colors, with remaining nodes shown in gray.}
    \label{fig:networks}
\end{figure*}

To assess whether the detected app clusters depend on the data filtering choices, we vary each filter parameter independently while holding the others fixed and measure the Jaccard overlap of flagged clusters against the reference configuration (Appendix~\ref{app:case1}, Table~\ref{tab:case1_threshold_overlap}). 
Overall, the reference configuration reliably captures most of the same suspicious app clusters across a range of filter settings. 

We summarize the behaviors observed in the ten largest components in Table~\ref{tab:case1_components} and report on these behaviors below. 
To characterize some clusters, we extract the top ten hashtags used by accounts within it, ranked by frequency and translated into English where needed. 
This provides a lightweight proxy for the topics and narratives promoted by a cluster without requiring a more complex topic model. 

\begin{table*}[t]
    \centering
    \setlength{\tabcolsep}{3pt}
    \renewcommand{\arraystretch}{1.1}
    \caption{Summary of Case~1 components (top-1\% similarity in app name similarity network). This and the next two tables have the following column names. \textbf{C}: component ID; \textbf{\#Apps}: number of apps; \textbf{\#U}: unique users; \textbf{RT\%}: retweet percentage; \textbf{$\langle r \rangle$}: average Pearson correlation of daily activity (higher = more synchronized). Country codes: MX=Mexico, VE=Venezuela, CAT=Catalonia, CU=Cuba, IR=Iran, RU=Russia, SA=Saudi Arabia, RS=Serbia, TR=Turkey, CN=China, EG=Egypt, HN=Honduras, ID=Indonesia, AE=UAE, UG=Uganda, EC=Ecuador.}
    \label{tab:case1_components}
    \begin{tabular}{@{}c r r r r l l l@{}}
    \hline
    \textbf{C} & \textbf{\#Apps} & \textbf{\#U} & \textbf{RT\%} & \textbf{$\langle r \rangle$} & \textbf{Campaign(s)} & \textbf{Behavior} & \textbf{Sample App} \\
    \hline
    1  & 48 & 1,453 & 0.4  & 0.02  & MX, VE, CAT, CU, IR, RU, SA, RS, TR, CN, EG, HN & API as a service     & \textit{erased12052391} \\
    2  & 33 & 33    & 0.0  & 0.00  & CN, HN, ID, IR, RU, SA, TR, VE                   & API permission abuse & \textit{RoundYearF 54}\\
    3  & 16 & 1,055 & 83.8 & 0.00  & SA                                               & Spoofed apps         & \textit{Twitter for I phone 0} \\
    4  & 8  & 36    & 0.0  & 0.1  & EG, SA                                           & Religious camouflage & \textit{Quranic App} \\
    5  & 6  & 203   & 9.0  & 0.04  & CN, EG, ID, SA, TR, RU                           & Follow-growth        & \textit{twitterrific} \\
    6  & 5  & 33    & 0.0  & 0.25  & RU                                               & Narrative promotion  & \textit{get-tweets.info07} \\
    7  & 5  & 77    & 0.0  & 0.10  & SA                                               & Spoofed apps         & \textit{twitter for apps ios} \\
    8  & 5  & 224   & 55.0 & 0.01  & CN, IR, CU, RS, UG, TR, SA, RU                   & Spoofed apps         & \textit{Tweepr on iOS} \\
    9  & 4  & 187   & 20.8 & -0.06 & CN, IR, RU, SA, AE, VE, TR, EG                   & Spoofed apps         & \textit{Tweetbot for iOS11}\\
    10 & 4  & 320   & 11.3 & 0.01  & CN, IR, CU, EG, EC, HN, ID, RU, VE, TR           & Follow-growth        & \textit{TweetCaster} \\
    \hline
    \end{tabular}
\end{table*}

\paragraph{\textbf{API as service:}}

The largest connected component consists of 48 apps, all with names beginning with \texttt{erased} followed by a variable number of digits. 
These apps were deployed in campaigns across many countries, the largest share being associated with Saudi Arabia.  
Some apps were used across multiple campaigns; for example, \texttt{erased138961} was used in the Catalonia, Honduras, Mexico and Cuba campaigns; \texttt{erased40502} was used in the Turkey, Indonesia, Iran, Saudi Arabia, Russia, China, Cuba, Egypt, Venezuela, Serbia and China campaigns. 
These apps primarily posted original tweets, with very few retweets, indicating they were used to generate rather than amplify content. Based on the top ten hashtags in each campaign, topics varied widely, including Vučić-related content in Serbia; religious content in Saudi Arabia, Egypt, and Iran; and follower growth in Turkey, China, Russia, and Venezuela. 
App usage also overlaps in time across campaigns. 
Taken together, the uniform naming pattern, cross-campaign reuse, temporal overlap, and emphasis on original content suggest that the \texttt{erased} apps were likely part of an API service that different entities rented to operate coordinated information operation accounts. 
We cannot confirm this directly as the apps are no longer accessible, but the same naming convention appearing across 12 unrelated campaigns simultaneously is unlikely to be coincidental.

\paragraph{\textbf{API permissions abuse:} }

The second connected component comprises 33 apps with different variation of the name \texttt{RoundYearFun}, e.g., 
\texttt{RoundYearFun57} and \texttt{RoundYearFun126}. 
These apps refer to a family of web-based games (\url{roundyearfun.org}). 
The content posted by accounts using these applications typically reflected game updates and was not politically oriented. 
However, we found various complaints by users who tried these games, reporting that the apps posted and followed other accounts on Twitter on their behalf without their acknowledgment. 
To investigate, we created a test account with minimal activity: one follower, following one account, and a single post. 
We authorized this account to interact with one of the \texttt{RoundYearFun} apps.
After the interaction, the app automatically followed three cryptocurrency-related accounts.
Additionally, the app liked four posts from distinct crypto-related accounts.
The test account did not receive any notifications for these actions unrelated to the game.
This suggests that the apps mislead users about their true purpose of promoting cryptocurrency accounts, and possibly engage in crypto-related fraud.
\texttt{RoundYearFun} games and apps are still active at writing time. 

\paragraph{\textbf{Spoofed apps:}}

Across multiple components, IO actors deliberately adopted app names that closely resemble legitimate Twitter apps, likely to obscure the true origin of their activity. 
In Components 3 and 7, a combined group of 21 apps includes spoofed apps such as \texttt{Twitter for I phone 0} and \texttt{Twitter for Android}, all linked to the Saudi Arabia campaign. 
This strategy extends beyond official apps. 
Component 8 and 9 includes apps that impersonate well-known third-party tools, namely \texttt{Tweepsmap}, a Twitter analytics platform, and \texttt{Tweepi}, a follower management service. 
One app name, \texttt{Tweetbot for iOS}, substitutes the Latin letter `O' with the visually identical Greek capital Omicron (Unicode U+039F) appearing identical to the legitimate app.
The repeated use of spoofed app names across multiple components highlights efforts to make activity appear as if it came from legitimate apps, making it harder to trace and detect.

\paragraph{\textbf{Religious camouflage:}}

Component 4 consists of eight apps with similar Arabic names that were active in the Saudi Arabia and Egypt campaigns. 
All associated activity involved original tweets with a religious theme. 
Each tweet included links to external websites, though most of these sites are no longer accessible for analysis; a few, such as \url{du3a.org}, remain active and continue to publish religious content. 
According to a report by the Stanford Internet Observatory~\cite{saudiArabia2019}, the religious content posted by the Saudi Arabia campaign was largely distributed through automated tweets, serving as a mask for activity aimed at promoting political messages and advancing geopolitical interests. 

\paragraph{\textbf{Follow-growth and narrative promotion:}}

Components 5 and 10 consist of legitimate third-party Twitter apps, while Component 6 used a custom apps (\texttt{get-tweets.info}). 
Despite different underlying tools, all three components combined follower growth along with promotion of domestic political topics.
In Component 5, top hashtags include generic follow-growth tactics such as \#followfriday and \#follow alongside Russian-language political hashtags, suggesting these apps were used for both audience building and Russian narrative promotion. 
Component 10 follows a similar pattern: accounts from Egypt, Cuba, and China used follow-back hashtags for audience building, while accounts from Turkey, Honduras, Ecuador, and Venezuela promoted domestic political narratives. 
In Component 6, accounts posted entirely original Russian cultural content, suggesting a focus on narrative promotion rather than audience building. 
Retweet ratios are very low across all three components, indicating a focus on original content production rather than amplification.
%

\subsection{Case 2: App network based on account overlap}

In the second case study, we model relationships between apps based on shared user accounts. 
We define the account overlap between apps $a_i$ and $a_j$ as:
\begin{equation}
    O(a_i, a_j) = 
    \frac{|A_i \cap A_j|}{\min(|A_i|, |A_j|) + 1},
\end{equation}
where $|A_i|$ is the total number of accounts using app $a_i$ and $|A_i \cap A_j|$ is the number of accounts common to both apps. 
Higher values of $O(a_i, a_j)$ indicate that more accounts use both apps, with $0 \leq O(a_i, a_j) \leq 1$. The constant 1 in the denominator prevents single-user apps from dominating the score.

Following the same threshold selection method as in Case 1, the percolation threshold occurs at the 96th percentile, and we again select the 99th percentile as our operating threshold, retaining only edges with an overlap value above 0.86. 
The resulting network contains 106 nodes and 141 edges across 20 components (Figure~\ref{fig:networks}B, Table~\ref{tab:network_stats} in Appendix). 
Sensitivity analysis (Table~\ref{tab:case2_threshold_overlap}, Appendix~\ref{app:case2}) confirms the core components are stable across parameter configurations. 
We extract the ten largest components, summarized in Table~\ref{tab:case2_components}, and describe the observed behaviors in detail below. 

\begin{table*}[t]
    \centering
    \caption{Summary of Case~2 components (top-1\% similarity in account-overlap network).
    Country codes: Q=Qatar, IRA=Internet Research Agency. (See previous table for additional codes.)
    }
    \label{tab:case2_components}
    \begin{tabular}{@{}c r r r r l l l@{}}
    \toprule
    \textbf{C} & \textbf{\#Apps} & \textbf{\#U} & \textbf{RT\%} & \textbf{$\langle r \rangle$} & \textbf{Campaign(s)} & \textbf{Behavior} & \textbf{Sample Apps} \\
    \midrule
    1  & 23 & 894 & 21.0  & 0.00  & EG, SA, Q & Religious content production      & \textit{Twitter for I~phone}, \textit{Dorar\_islamic} \\
    2  & 8  & 39  & 29.0  & -0.05 & RU(IRA)                        & No prominent behavior   & \textit{Projekt program for you 5}, \textit{mr.~Anderson} \\
    3  & 7  & 851 & 96.4  & 0.54  & RU(IRA),                 & Amplification (synchronous) & \textit{ToptwitRu}, \textit{rostislav} \\
    4  & 6  & 36  & 87.0  & -0.01  & SA              & Amplification               & \textit{14TUT14},  \textit{erased12218827} \\
    5  & 6  & 457 & 11.3  & 0.06  & TR, RU (bridged)   & Attacks on opposition party & \textit{Hootsuite}, \textit{RetweetHavuzu 5.9} \\
    6  & 5  & 28  & 82.2  & 0.17  & VE                  & Amplification (political)   & \textit{GranPoloPatriotico}, \textit{PSUV} \\
    7  & 5  & 12  & 100.0 & 0.61  & RU(IRA)                & Amplification (fake news)   & \textit{HelperT}, \textit{T-Support} \\
    8  & 4  & 456 & 86.3  & -0.01  & RU(IRA)                & Amplification              & \textit{newtwittersky}, \textit{ohwee}\\
    9  & 4  & 459 & 0.0   & 0.17    & EG, SA, Q & Religious content production      & \textit{Twetat.Com}, \textit{athkarApp} \\
    10 & 4  & 13  & 82.7  & -0.08 & SA               & Amplification               & \textit{Dsalguuero}, \textit{erased12279581}\\
    \bottomrule
    \end{tabular}
\end{table*}

\paragraph{\textbf{Religious content production:}}

Components~1 and~9 post predominantly original tweets (79\% and 100\%, respectively) with religious hashtags. They are active in the Egypt, Saudi Arabia, and Qatar campaigns. 
Cross-referencing with Case~1 reveals that 12 of Component~1's apps and one of Component~9's apps also appear in Case~1 components, indicating that these apps were also flagged by name similarity. 
Some campaigns can be detected using multiple app-based indicators of coordination. 

\begin{figure}
    \centering
    \includegraphics[width=\columnwidth]{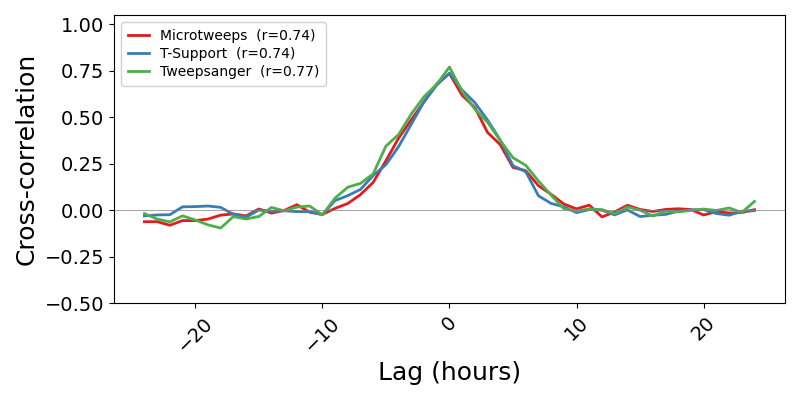}
    \caption{
    Cross-correlation of hourly retweet activity of \texttt{Microtweeps}, \texttt{T-Support}, and \texttt{Tweepsanger} against \texttt{HelperT} as the reference app in Component~7 (Case~2). 
    }
    \Description{
    Cross-correlation of hourly retweet activity of \texttt{Microtweeps}, \texttt{T-Support}, and \texttt{Tweepsanger} against \texttt{HelperT} as the reference app in Component~7 (Case~2).
    }
    \label{fig:comp_6}
\end{figure}

\paragraph{\textbf{Amplification:}}

Amplification is the most commonly observed behavior in this case, spanning multiple campaigns. 
Several campaigns detected by high account overlap exhibit high retweet ratios, between 82--100\%. 
Components~4 and 10 are associated with the Saudi Arabia campaign. 
They amplified accounts that are now suspended, but not accounts in Components~1 or 9 that were affiliated with the same campaign. 
Component~6 contains five politically branded apps from Venezuela (\texttt{GranPoloPatriotico}, \texttt{RedTuiteros}, \texttt{PSUV}, \texttt{PatriaGrande}, and \texttt{OlaBolivariana}). 
A Stanford report~\cite{venezuela} documents their use in propagating identical tweets and retweets supporting Hugo Chávez and later Nicolás Maduro. 
The \texttt{PSUV} app is directly tied to the United Socialist Party of Venezuela. 
Components~3 and 8 amplify content from the Russia IRA campaign.
Component~7 stands out: its five apps (\texttt{Jerusalem}, \texttt{Microtweeps}, \texttt{HelperT}, \texttt{T-Support}, \texttt{Tweepanger}) operate through 12 unique accounts with an average overlap of nine accounts. 
Let us define pairwise synchronicity as the Pearson correlation $r$ in daily activity between two apps. 
We measure a high average synchronicity ($\langle r \rangle = 0.61$) across pairs of apps in this component.  
Further analysis reveals that four of the apps were active in an overlapping period of time and exhibit high synchronicity. 
Figure~\ref{fig:comp_6} shows that the cross-correlation among posts by these four apps peaks at lag zero ($r \approx 0.75$) at an hourly resolution, meaning that their activities rise and fall at exactly the same times.  
All activity consists of retweets targeting a mix of legitimate news sources (\textit{@Independent}, \textit{@thehill}, \textit{@TIME}, \textit{@guardian}) alongside now-suspended fake news accounts (e.g., \textit{@NYC\_Everyday}, \textit{@TodayPittsburgh}, \textit{@Seattle\_Post}, \textit{@WashDCOnline}, \textit{@KansasCityDNews}, \textit{@ChicagoILDaily}, \textit{@NOrleansDaily}, \textit{@PhoenixNewsAZ}, \textit{@DetroitPost}). 
This pattern suggests a single entity using multiple apps to control multiple accounts that amplify fake news. 

\paragraph{\textbf{Attacks on opposition party:}}

Component~5 contains two sub-clusters bridged by a commercial platform. 
Turkish apps \linebreak \texttt{RetweetHavuzu~5.9}, \texttt{RetweetHavuzu~VOL2.1}, and \linebreak \texttt{Blaq for PlayBook\texttrademark} share seven accounts, while Russian apps \texttt{get-tweets.info5} and \texttt{get-tweets.info07} share 8--10 accounts. 
These sub-clusters have no direct overlap with each other; they appear in the same component only because both share users with \texttt{Hootsuite}. 
Based on the top hashtags, both sub-clusters show a mix of generic tweets and attacks on opposition parties --- the Republican People's Party in the Turkish sub-cluster, Fair Russia and Navalny-aligned opposition in the Russian sub-cluster.

\paragraph{\textbf{No prominent behavior:}}

Component~2 contains apps with non-descriptive or deliberately obfuscated names (e.g., \texttt{Projekt program for you 5}, \texttt{mr.~Anderson}, \texttt{break\_bed\_boy}) operated by 39 IRA accounts.
It has moderate retweet ratio (29\%), apolitical lifestyle hashtags (sports, music, news, health), and lack of app synchronicity.
We could not identify any prominent manipulation tactic associated with this component.

\subsection{Case 3: Account network based on app usage}

In the third case study, we construct a coordination network in which nodes represent accounts and edges represent similarity in app usage patterns among them.
We follow the approach of~\citet{pacheco2021uncovering} to measure app usage similarity, augmented with the percolation-based threshold selection introduced in Case~1 to determine a data-informed edge cutoff. 
The procedure consists of three steps for any given indicator:

\begin{enumerate}
    \item \textbf{Bipartite graph construction.} We extract coordination indicators together with their associated user accounts and build a bipartite graph: one set of nodes represents user accounts, the other represents indicators (apps in this particular case). 
    We use TF-IDF weights to measure the importance of an indicator in an account's usage pattern.

    \item \textbf{Account–account network projection.} The bipartite graph is projected onto an account–account similarity network, where an edge weight is the cosine similarity between the TF-IDF vectors of indicators of the corresponding pair of accounts.

    \item \textbf{Filtering.} We rank edges by similarity and apply percolation-based threshold selection to identify the structural transition point. 
    In the app indicator case, the number of components and the size of the largest component stabilize beyond the 99th percentile (Appendix~\ref{app:case3}, Figure~\ref{fig:percolation_analysis_case3}).  
    We select the 99.9th percentile, corresponding to a cosine similarity threshold of one, retaining only accounts that share identical app usage patterns. 
\end{enumerate}

The resulting account network in this case contains 1,747 nodes and 55,039 edges across 91 components (Figure~\ref{fig:networks}C, (Table~\ref{tab:network_stats} in Appendix). 
Sensitivity analysis (Table~\ref{tab:case3_threshold_overlap}, Appendix~\ref{app:case3}) shows robust clusters across parameter settings.
The common behavioral pattern observed in the top 10 largest components are summarized in Table~\ref{tab:case3_components} and described in detail below.

\begin{table*}[t]
    \centering
    \caption{Summary of Case~3 components (top-0.1\% similarity on user similarity network). See previous table for country codes.
    }
    \label{tab:case3_components}
    \begin{tabular}{@{}c r r r l l l@{}}
    \toprule
    \textbf{C} & \textbf{\#U} & \textbf{RT\%} & \textbf{$\langle r \rangle$} & \textbf{Campaign(s)} & \textbf{Behavior} & \textbf{Sample App} \\
    \midrule
    1  & 260 & 54.21 & 0.38 & EC                             & Synchronized posting & \textit{MoreSocial.net} \\
    2  & 139 & 27.91 & 0.25 & RU(IRA)                            & Propoganda and fake news                & \textit{masss post5} \\
    3  & 118 &  3.15 & 0.12 & ID, CN                         & Posting through a bot platform                & \textit{twittbot.net} \\
    4  & 105 &  2.85 & 0.19 & RU(IRA), VE                        & Propaganda and fake news    & \textit{Twibble.io} \\
    5  &  94 & 40.99 & 0.08 & TR, IR, SA, VE, ID, CN, CU, HN, EC & Miscellaneous & \textit{TweetCaster} \\
    6  &  84 & 42.97 & 0.05 & RU(IRA), IR                        & Miscellaneous            & \textit{Twitter Lite} \\
    7  &  81 & 93.44 & 0.11 & CN, EG, TR                     & Amplification                   & \textit{Twitter for Google TV} \\
    8  &  79 & 35.26 & 0.38 & UAE                            & Synchronized posting & \textit{SocialReport.com} \\
    9  &  69 & 91.49 & 0.10 & TR, SA, CN, IR                 & Amplification                   & \textit{Twitter for Mac} \\
    10 &  58 & 11.94 & 0.25 & RU(IRA)                            & Propoganda and fake news                & \textit{masss post4} \\
    \bottomrule
    \end{tabular}
\end{table*}

\paragraph{\textbf{Synchronized posting via social media management tools:}}

Components~1 (Ecuador) and 8 (UAE) exhibit coordination driven by social media management tools. 
In Component~1, the average synchronicity is $\langle r \rangle = 0.38$, with  5810 out of 33,670 user pairs having $r > 0.8$, indicating strongly coordinated posting. 
Content combines political commentary (\#RodasElPeorAlcalde), sports (\#RealMadrid), and commercial promotions (\#SuperOfertas); a persona diversification strategy designed to appear organic. 
This is consistent with a report by the \citet{aspi_ecuador}, which identified the same management tool as supporting pro-government narratives during Ecuadorian President Lenín Moreno's administration. Component~8 exhibits a similar pattern with an average synchronicity of $\langle r \rangle = 0.38$ and 150 out of 3,081 user pairs exceeding 0.8 correlation.
Each account posts in almost nine different languages on average, suggesting systematic targeting of a multilingual audience across regional geopolitical topics including Qatar, Somalia, Yemen, and the Gulf World Cup.

\paragraph{\textbf{Propaganda and fake news dissemination:}}
Components~2, 4, and 10 disseminate state-aligned content at scale. 
Components~2 and 10 are both IRA clusters that posted through custom tools with versioned names (\texttt{masss~post4}, \texttt{masss~post5}), suggesting iterative development of internal posting infrastructure. 
Both components amplify known disinformation narratives, including the fabricated Columbian Chemicals incident and false Fukushima narratives~\cite{chen2015agency}, reflected in shared hashtags such as \#ColumbianChemicals and \#FukushimaAgain. 
Despite amplifying the same narratives, the two components share no users, suggesting the operation was deliberately split across separate account sets to reduce the risk of detection. 
Component~4 bridges the IRA and Venezuela campaigns through shared RSS-to-Twitter automation (\texttt{Twibble.io}). 
On the IRA side, accounts disseminate articles from state-funded outlets such as \textit{actualidad.rt.com} and \textit{fedpress.ru}, with \textit{actualidad.rt.com} documented as a vehicle for Kremlin propaganda~\cite{espreso2024rt_cloning}. 
On the Venezuelan side, hashtags such as \#LegionFANB and \#SomosPatriaChavista promote support for the Bolivarian Armed Forces and pro-Chavista ideology, consistent with documented state-linked amplification campaigns on Venezuelan Twitter~\cite{dfrlab_venezuela_trolltracker}.
The fact that two unrelated state-aligned campaigns independently used the same tool to automate their posting shows how a single commercial platform can end up linking otherwise separate operations.

\paragraph{\textbf{Posting through a bot platform:}}
Component 3 captures coordinated activity carried out through a bot platform. 
\texttt{Twittbot.net} is a web-based service that enables users to create and manage automated bots. 
In the China-related campaign, the content combines entertainment themes such as K-pop with narratives targeting Guo Wengui, a known focus of coordinated influence operations attributed to Chinese state-linked actors~\cite{sio2021china}. 
In the Indonesia-related campaign, the content primarily centers on the Free West Papua movement.

\paragraph{\textbf{Amplification:}}
Components~7 and 9 are primarily amplifiers, with retweet ratios above 90\%. 
Component~7 spans three campaigns. China-related activity centers on follow-back engagement and Egypt-related content is adult-oriented. We did not find consistent themes in Turkey-related content. 
Component~9 includes narratives related to the Hong Kong protests (China campaign), the Yemen conflict (Saudi Arabia campaign), and a policy debate related to teacher appointments (Turkey campaign). Surprisingly, accounts in this component related to the Iran campaign amplify messages with anti-regime slogans, as indicated by hashtags that translate to \#there\_is\_no\_compulsory\_hijab and \#i\_am\_an\_overthrower. 

\paragraph{\textbf{Miscellaneous:}}
Components~5 and 6 display a mix of behaviors.
Each campaign in Component~5 pursues distinct objectives, ranging from Turkish employment activism (\#5bingidamuhendisi) to Iranian anti-Israel narratives and Venezuelan amplification of Russian state media. 
Component~6 links IRA and Iranian operations.
IRA accounts primarily produce original pro-Russia content (\#ImageOfRussia, \#Putin, \#Syria), while Iranian accounts act as amplifiers (62--68\% retweet ratios) promoting anti-Israel and pro-Palestinian narratives (\#FreePalestine, \#DeleteIsrael, \#QudsDay). Both campaigns consistently amplify state-affiliated media sources: \textit{rg.ru} and \textit{almasdarnews.com} for the IRA, and \textit{en.farsnews.com} for Iran.

\section{App usage as a coordination indicator}

To answer RQ2, we test whether adding app usage as a coordination indicator improves the detection of coordinated accounts compared to using other indicators alone. 
For this analysis, we use a labeled dataset of 26  campaigns that include both IO and control accounts (Section~\ref{sec:labeled-datasets}). 
Since coordination tactics vary substantially across operations, supervised models trained on one campaign may not generalize reliably to others. 
Therefore we use an unsupervised network-based detection method that operates at the multiple-indicator level, linking accounts based on shared behavioral signals.

\subsection{Indicators}

We compare two setups: one using five established indicators (shared hashtags, shared URLs, retweeted users, retweeted posts, and posting synchronicity) and one that adds app usage. 
In addition to the filtering explained in Section~\ref{sec:labeled-datasets}, for posting synchronicity we bin tweets into 30-minute intervals~\cite{pacheco2021uncovering}. 
We construct user similarity networks following Steps~(1) and (2) of Case~3. 
These processes result in separate similarity networks, one per indicator, which are combined to classify nodes as IO or control.

\subsection{Maximum indicator support}

We are aware of two state-of-the-art unsupervised methods operating at the multiple-indicator level: the node-pruning approach proposed by \citet{luceri2024} and the time-aware collaboration method of \citet{time_aware}. 
The time-aware collaboration method makes a key assumption that coordination occurs within short bursts of activity. 
Therefore temporal information is folded directly into each indicator's similarity computation. 
This assumption does not hold in general, as accounts that are part of the same campaign may act at different times, and we wish to detect such coordination even when occurring without synchronicity. 
Although synchronicity is not a necessary condition for coordination, it may be a sufficient condition; that is the reason for our use of posting synchronicity as a separate indicator. 

The node-pruning method of \citet{luceri2024} combines all indicator similarity networks using a union approach: a single edge is retained between any pair of accounts connected in at least one indicator network.
Unweighted eigenvector centrality is then computed for each node in the fused network, and nodes are ranked by centrality. 
Detection performance is evaluated by sweeping centrality percentile thresholds. 
This method assumes that every edge with non-zero similarity is equally meaningful. 
This is not the case in our dataset, and fused networks are therefore very dense --- the minimum density is 0.6 across campaign networks. 
This makes centrality a weak indicator of coordination, as illustrated in Figure~\ref{fig:luceri}, Appendix~\ref{app:luceri}. 

To address the above limitations, we combine the different indicator networks using an assumption-free aggregation method we call \textit{maximum indicator support}. 
As the similarities in different indicator networks are not directly comparable, we need a statistical way to filter them before the networks can be combined.
For each indicator network, we apply a permutation-based null model to calculate the $p$-value of each edge.  
One thousand permutations are obtained by shuffling edges in the bipartite network while preserving degrees of both account and indicator nodes. 
Similarities are then computed for each permutation of the network. 
The $p$-value of each edge in the indicator network can then be calculated as the number of permutations leading to equal or larger similarity values between the two nodes connected by that edge across network permutations.
The $p$-values of edges obtained in this way can be compared across indicator networks, and a single threshold can be applied to remove low-similarity edges ($p \ge 0.05$ after Benjamini–Hochberg correction).  
This calibration step is particularly important for the app indicator, where popular apps create high incidental similarity that would otherwise inflate coordination signals.
We can finally combine the networks: for each pair of accounts, we count the number of indicator networks in which they are still connected by a significant edge. 
This count is aggregated to the node level by taking the maximum across all of a node's edges (\textit{maximum indicator support}), reflecting the highest number of indicators that simultaneously flag any single relationship involving that account.

\subsection{Results}

IO accounts show a significantly higher maximum indicator support score than control accounts in 16 out of 26 campaigns (Mann-Whitney U test, $p < 0.05$) when all indicators are considered (Figure~\ref{fig:man_whitney_all} in Appendix~\ref{app:coordination_indicator}). 
This suggests that IO accounts tend to leave traces across more behavioral dimensions. 

\begin{figure}
    \centering
    \includegraphics[width=0.75\columnwidth]{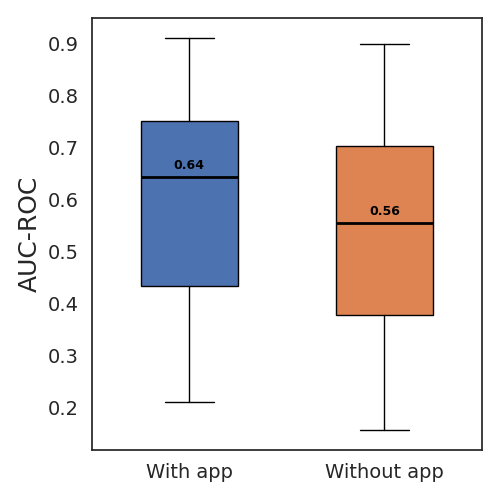}
    \caption{AUC-ROC distributions with and without the app indicator across 26 campaigns using maximum indicator support. 
    A Wilcoxon paired test confirms that detection performance is significantly higher when the app indicator is included ($p<0.05$).}
     \Description{AUC-ROC distributions with and without the app indicator across 26 campaigns using maximum indicator support. 
    A Wilcoxon paired test confirms that detection performance is significantly higher when the app indicator is included ($p<0.0001$).}
    \label{fig:max_indicator_results}
\end{figure}

We evaluate whether this separation translates into detection performance by using the maximum indicator support score as a ranking signal. 
As shown in Figure~\ref{fig:max_indicator_results}, including the app indicator yields a moderate but consistent improvement across campaigns with median AUC-ROC increasing from 0.56 to 0.64.
This difference is statistically significant according to a Wilcoxon signed-rank test ($p<0.0001$).
A campaign-level breakdown further confirms that this improvement is consistent across campaigns, with the single exception of the Cuba campaign (Figure~\ref{fig:auc_roc_wilcoxon} in Appendix~\ref{app:coordination_indicator}). 

\section{Discussion}

APIs and third-party app access serve as central gateways into social media ecosystems. 
Regardless of whether content is human- or AI-generated, it is the ability to automate activity through these interfaces that enables manipulation at scale. 
App-level infrastructure offers multiple strategic advantages for influence operators: decreased human costs of managing many accounts manually, distribution of activity across different apps to circumvent platform rate limits, and reduction in the likelihood of triggering automation detection systems. 
Moreover, API tokens granted to third-party apps can remain active even after account suspension, allowing coordinated activity to persist or reemerge through 
alternate accounts.

The three case studies demonstrate that third-party applications constitute a critical yet underexplored infrastructure for social media manipulation.
Case~1 reveals how app naming patterns expose shared infrastructure: API-as-a-service providers operating across multiple campaigns, gaming apps that perform unauthorized engagement actions without user knowledge, and spoofed app names designed to obscure the origin of activity.
Case~2 shows that apps sharing overlapping user accounts expose centrally controlled amplification networks, where the same pool of accounts posts through multiple apps to boost content at scale.
Case~3 shows that accounts sharing similar app usage patterns coordinate through distinct tactics, including synchronized social media activity and custom fake news sources.
Adding app usage as a coordination indicator consistently improves detection across all campaigns, though the improvement is moderate.

The findings demonstrate that key traces of coordination are embedded within the technological mechanisms used to deploy campaigns, not just in their visible content. 
Because the same apps appear across distinct campaigns, app-level analysis provides a scalable signal for detection and attribution.

Platforms often hide how apps access platform features. In particular, Twitter/X has not only made API access extremely expensive for researchers, but has also removed access to app metadata entirely, making this class of analysis impossible going forward.
Increasing transparency around app-level metadata can improve governance, strengthen enforcement, and promote accountability among third-party developers. 


Our study has several limitations. 
    The case studies focus on historical datasets from specific time frames, which may not capture evolving tactics or newly emerging applications used in current operations.
    We applied fixed thresholds to filter and detect coordination. Alternative threshold values might yield slightly different detection outcomes.
    Content similarity was not used as a coordination indicator in Case Study~3 because the data spans long periods and many diverse topics, reducing the effectiveness of semantic similarity measures. 
    Finally, the present analyses are limited to a platform (Twitter) and time (before 2023) when IO datasets and app metadata were accessible. Our results highlight the analytical and transparency value of such metadata, emphasizing the need for emerging platforms to expose and preserve similar information.

\subsection*{Acknowledgements}

This work was supported in part by Knight Foundation, DARPA (contract HR001121C0169), and Lilly Endowment, Inc., through its support for the Indiana University Pervasive Technology Institute.

\bibliographystyle{ACM-Reference-Format}
\bibliography{ref}

\newpage
\appendix
\section{Appendix}

\begin{table}[H]
    \centering
    \footnotesize
    \caption{Network statistics across case studies}
    \label{tab:network_stats}
    \begin{tabular}{ccccccc}
    \hline
    \textbf{Case} & \multicolumn{2}{c}{\textbf{Threshold}} & \textbf{Nodes} & \textbf{Edges} & \textbf{Density} & \textbf{Components} \\
    & \textbf{\%ile} & \textbf{Similarity} & & & &  \\
    \hline
    1 & 99 & 0.48 & 203 & 1228 & 0.06 & 41 \\
    2 & 99  & 0.86  &  106  & 141   &  0.025  & 20  \\
    3 &  99.9  &  1  & 1747   &  55039   &  0.03     & 91 \\
    \hline
    \end{tabular}
\end{table}

\subsection{Case study 1}
\label{app:case1}

\begin{figure}[H]
    \centering
    \includegraphics[width=\columnwidth]{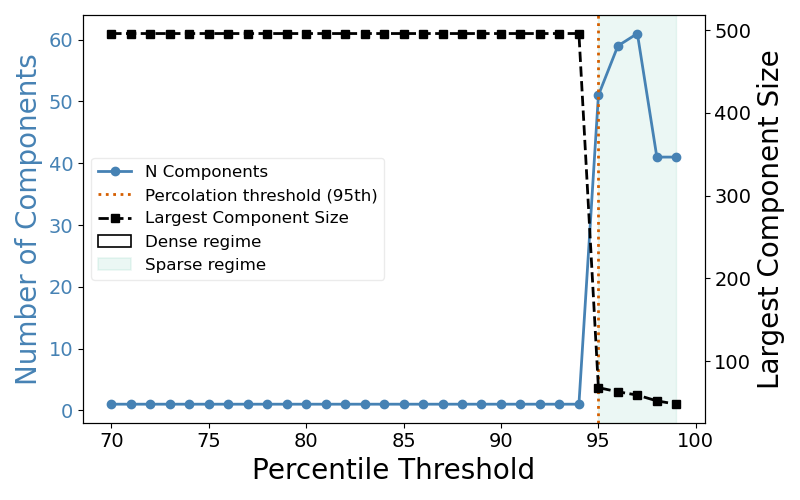}
    \caption{Edge-percolation analysis for Case~1 (app name similarity). The left axis shows the number of connected components and the right axis shows the largest component size, as a function of similarity percentile threshold. The dashed line marks the 95th percentile threshold, beyond which the network decomposes into distinct app clusters.}
    \Description{Edge-percolation analysis for Case~1 (app name similarity). The left axis shows the number of connected components and the right axis shows the largest component size, as a function of similarity percentile threshold. The dashed line marks the 95th percentile threshold, beyond which the network decomposes into distinct app clusters.
    }
    \label{fig:percolation_analysis_case1}
\end{figure}

\begin{table}[H]
\centering
\caption{
    \textbf{Case study 1: Parameter sensitivity analysis.} 
    Four parameters together define the filtering thresholds configuration: \texttt{tpa} (tweets per app), \texttt{tpu} (tweets per user), \texttt{upa$_{\min}$} (minimum users per app), and \texttt{upa}$_{\max}$ (maximum users per app). 
    Each row varies one parameter while holding the others fixed. 
    We report the Jaccard similarity between the top-1\% apps identified using each configuration versus the reference configuration (\texttt{tpa=100, tpu=10, upa$_{\min}$=5, upa$_{\max}$=500}). 
    High Jaccard values indicate that the same app pairs dominate the top-1\% regardless of the filtering choice. 
    The components are stable across \texttt{tpa}, \texttt{tpu}, and \texttt{ upa$_{\max}$} (Jaccard $\geq$ 0.89), while \texttt{upa$_{\min}$} shows greater sensitivity
    (0.68--0.77).
}
\begin{tabular}{llr}
\hline
\textbf{Parameter} & \textbf{Value} & \textbf{Jaccard} \\
\hline
\multirow{2}{*}{tpa}          & 50   & 0.91 \\
                               & 200  & 0.89 \\
\hline
\multirow{2}{*}{tpu}          & 5    & 0.99 \\
                               & 50   & 0.98 \\
\hline
\multirow{2}{*}{upa$_{\min}$} & 3    & 0.77 \\
                               & 10   & 0.68 \\
\hline
\multirow{2}{*}{upa$_{\max}$} & 300  & 0.94 \\
                               & 1000 & 0.96 \\
\hline
\end{tabular}
\label{tab:case1_threshold_overlap}
\end{table}

\subsection{Case study 2}
\label{app:case2}

\begin{figure}[H]
    \centering
    \includegraphics[width=\columnwidth]{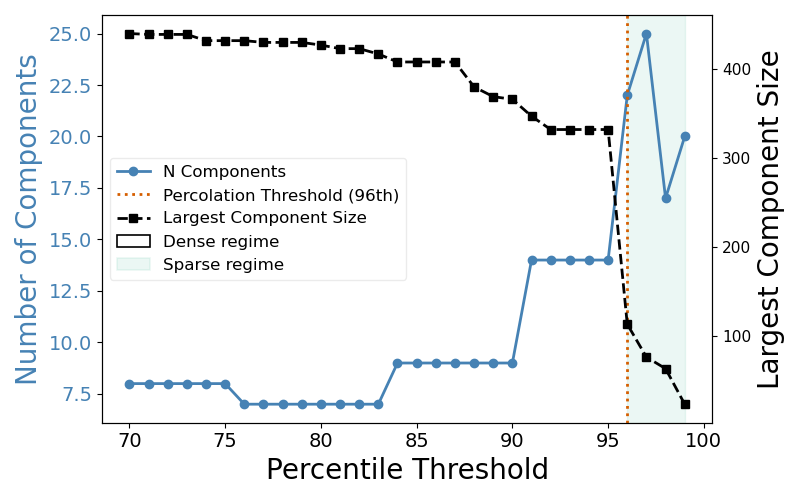}
     \caption{Edge-percolation analysis for Case~2 (apps similarity based on account overlap). The left axis shows the number of connected components and the right axis shows the largest component size, as a function of similarity percentile threshold. The dashed line marks the 96th percentile threshold, beyond which the network decomposes into distinct app clusters.}
    \Description{
    Edge-percolation analysis for Case~2 (apps similarity based on account overlap). The left axis shows the number of connected components and the right axis shows the largest component size, as a function of similarity percentile threshold. The dashed line marks the 96th percentile threshold, beyond which the network decomposes into distinct app clusters.
    }
    \label{fig:percolation_analysis_case2}
\end{figure}

\begin{table}[H]
\centering
\small
\caption{
    \textbf{Case study 2: Parameter sensitivity analysis.}
    Four parameters together define the filtering thresholds configuration: \texttt{tpa} (tweets per app), \texttt{tpu} (tweets per user), \texttt{upa$_{\min}$} (minimum users per app), and \texttt{upa}$_{\max}$ (maximum users per app). 
    Each row varies one parameter while holding the others fixed. 
    We report the Jaccard similarity between the top-1\% apps identified using each configuration versus the reference configuration (\texttt{tpa=100, tpu=10, upa$_{\min}$=5, upa$_{\max}$=500}). 
    High Jaccard values indicate that the same app pairs dominate the top-1\% regardless of the filtering choice. 
    The components are stable across \texttt{tpu} (Jaccard $\geq$ 0.99), moderately stable across \texttt{tpa} (0.79--0.91) both \texttt{upa$_{\min}$} and \texttt{upa$_{\max}$} show greater sensitivity (0.66--0.91 and 0.70--0.76, respectively), suggesting that which apps are included are determined by how popular the apps are.
}
\label{tab:case2_threshold_overlap}
\begin{tabular}{llr}
\toprule
\textbf{Parameter} & \textbf{Value} & \textbf{Jaccard} \\
\midrule
\multirow{2}{*}{tpa}          & 50   & 0.91 \\
                               & 200  & 0.79 \\
\midrule
\multirow{2}{*}{tpu}          & 5    & 1.00 \\
                               & 50   & 0.99 \\
\midrule
\multirow{2}{*}{upa$_{\min}$} & 3    & 0.91 \\
                               & 10   & 0.66 \\
\midrule
\multirow{2}{*}{upa$_{\max}$} & 300  & 0.70 \\
                               & 1000 & 0.76 \\
\bottomrule
\end{tabular}
\end{table}

\subsection{Case study 3}
\label{app:case3}

\begin{figure}[H]
    \centering
    \includegraphics[width=\columnwidth]{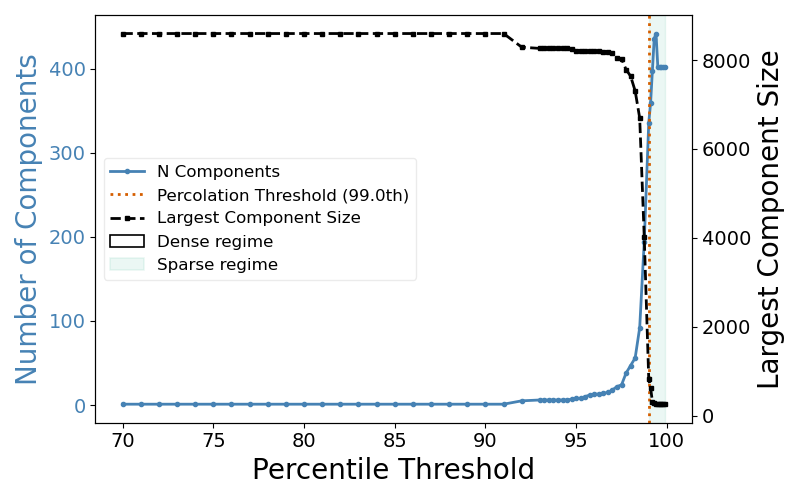}
   \caption{
   Edge-percolation analysis for Case~3 (user similarity based on app usage). The left axis shows the number of connected components and the right axis shows the largest component size, as a function of similarity percentile threshold. The dashed line marks the 99th percentile threshold, beyond which the network decomposes into distinct app clusters.
   }
   \Description{Edge-percolation analysis for Case~3 (user similarity based on app usage). The left axis shows the number of connected components and the right axis shows the largest component size, as a function of similarity percentile threshold. The dashed line marks the 99th percentile threshold, beyond which the network decomposes into distinct app clusters.}
    \label{fig:percolation_analysis_case3}
\end{figure}

\begin{table}[H]
\centering
\caption{
    \textbf{Case study 3: Parameter sensitivity analysis.}
    Four parameters together define the filtering thresholds configuration: \texttt{tpa} (tweets per app), \texttt{tpu} (tweets per user), \texttt{upa$_{\min}$} (minimum users per app), and \texttt{upa}$_{\max}$ (maximum users per app). 
    Each row varies one parameter while holding the others fixed. 
    We report the Jaccard similarity between the top-0.1\% users identified using each configuration versus the reference configuration (\texttt{tpa=100, tpu=10, upa$_{\min}$=5, upa$_{\max}$=500}). 
    High Jaccard values indicate that the same users pairs dominate the top-0.1\% regardless of the filtering choice. 
    The components are highly stable across \texttt{tpa}, \texttt{tpu}, and \texttt{upa$_{\min}$}
    (Jaccard $\geq$ 0.91), while \texttt{upa$_{\max}$} shows the greatest sensitivity (0.63--0.75), suggesting that the highly popular apps have strongest influence on which users appear in the top-0.1\%.
}
\label{tab:case3_threshold_overlap}
\begin{tabular}{llr}
\toprule
\textbf{Parameter} & \textbf{Value} & \textbf{Jaccard} \\
\midrule
\multirow{2}{*}{tpa}          & 50   & 0.98 \\
                               & 200  & 0.97 \\
\midrule
\multirow{2}{*}{tpu}          & 5    & 0.97 \\
                               & 50   & 0.91 \\
\midrule
\multirow{2}{*}{upa$_{\min}$} & 3    & 0.97 \\
                               & 10   & 0.96 \\
\midrule
\multirow{2}{*}{upa$_{\max}$} & 300  & 0.75 \\
                               & 1000 & 0.63 \\
\bottomrule
\end{tabular}
\end{table}

\subsection{Node-pruning method results}
\label{app:luceri}

\begin{figure}[H]
    \centering
    \includegraphics[width=0.5\linewidth]{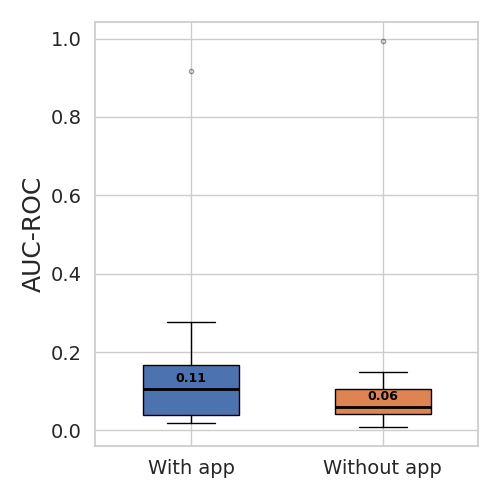}
    \caption{Result based on the node-pruning method by \citet{luceri2024}.}
    \Description{Result based on the node-pruning method by \citet{luceri2024}.}
    \label{fig:luceri}
\end{figure}

\subsection{App usage as coordination indicator}
\label{app:coordination_indicator}

\begin{figure}[H]
    \centering
    \includegraphics[width=\columnwidth]{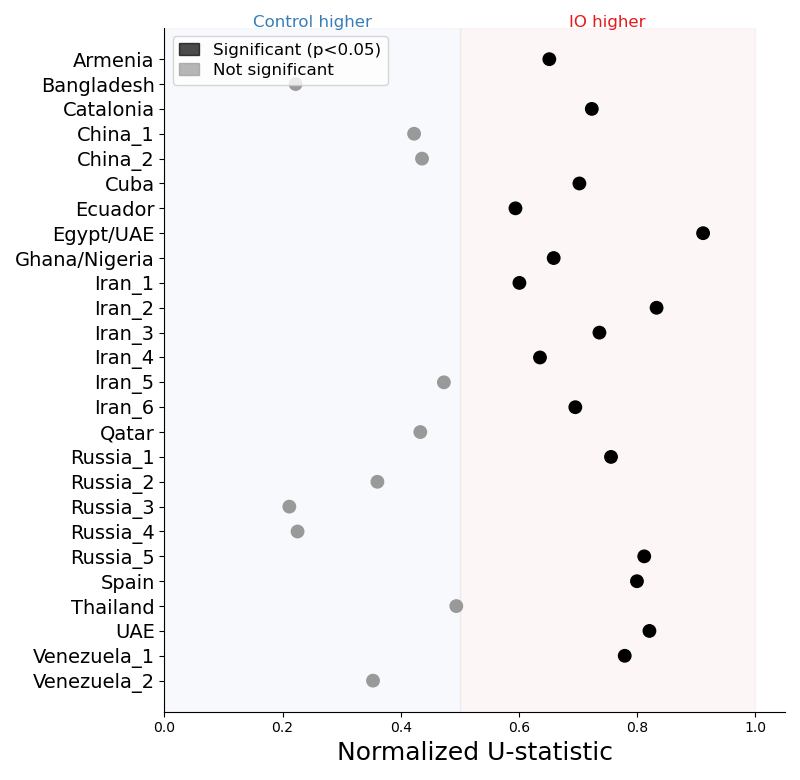}
    \caption{
    Normalized Mann-Whitney $\hat{U}$ statistic ($\hat{U} = U / (n_{\text{IO}} \times n_{\text{control}})$) comparing maximum indicator support scores $X$ between IO and control accounts across 26 campaigns, with the app indicator. A one-sided test is applied under the alternative hypothesis $H_1: P(X_{\text{IO}} > X_{\text{control}}) > 0.5$. $\hat{U} > 0.5$ indicates that IO accounts tend to score higher than control accounts, $\hat{U} = 0.5$ indicates no systematic difference, and $\hat{U} < 0.5$ indicates that control accounts tend to score higher. Black dots denote campaigns where the hypothesis is supported after Benjamini-Hochberg correction ($p < 0.05$).
    }
    \Description{
    Normalized Mann-Whitney $\hat{U}$ statistic ($\hat{U} = U / (n_{\text{IO}} \times n_{\text{control}})$) comparing maximum indicator support scores $X$ between IO and control accounts across 26 campaigns, with the app indicator. A one-sided test is applied under the alternative hypothesis $H_1: P(X_{\text{IO}} > X_{\text{control}}) > 0.5$. $\hat{U} > 0.5$ indicates that IO accounts tend to score higher than control accounts, $\hat{U} = 0.5$ indicates no systematic difference, and $\hat{U} < 0.5$ indicates that control accounts tend to score higher. Black dots denote campaigns where the hypothesis is supported after Benjamini-Hochberg correction ($p < 0.05$).
    }
    \label{fig:man_whitney_all}
\end{figure}

\begin{figure}[H]
    \centering
    \includegraphics[width=\columnwidth]{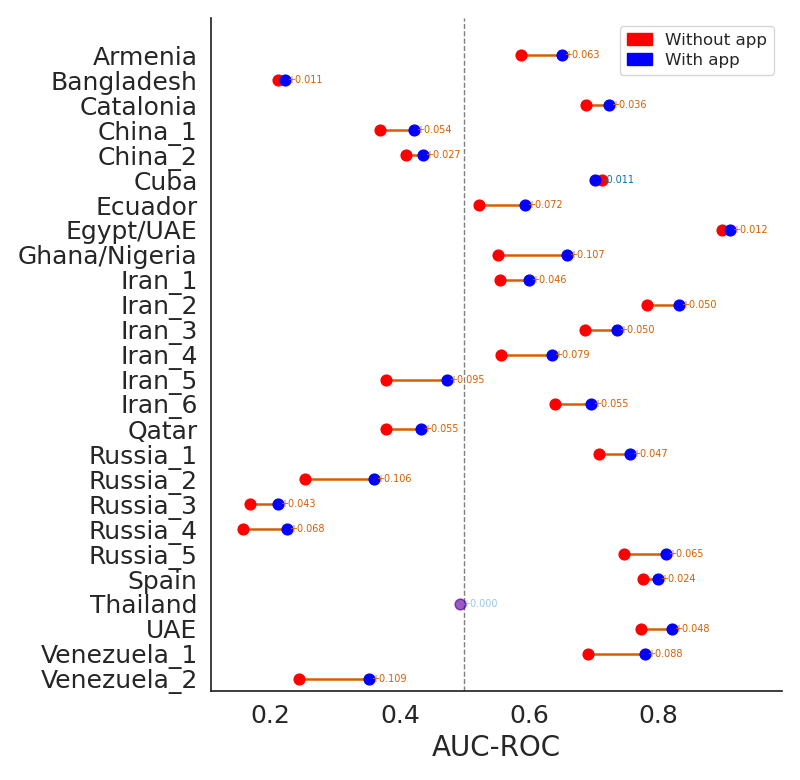
    }
    \caption{
    Per-campaign change in AUC-ROC when the app indicator is included compared to excluded, using maximum indicator support across 26 campaigns. Each line connects the AUC-ROC of a single campaign without the app indicator (red) to that of the same campaign with the app indicator (blue). A rightward slope indicates an improvement in detection performance. 
    }
    \Description{
    Per-campaign change in AUC-ROC when the app indicator is included compared to excluded, using maximum indicator support across 26 campaigns. Each line connects the AUC-ROC of a single campaign without the app indicator (red) to that of the same campaign with the app indicator (blue). A rightward slope indicates an improvement in detection performance. 
    }
    \label{fig:auc_roc_wilcoxon}
\end{figure}

\end{document}